\documentclass[twocolumn,aps,prb,floats,showpacs]{revtex4}
\newcommand{\beq}{\begin{eqnarray}}
\newcommand{\eeq}{\end{eqnarray}}

\newcommand{\bs}{\begin{subequations}}
\newcommand{\es}{\end{subequations}}

\usepackage{epsfig}
\usepackage{dcolumn}
\usepackage{bm}
\usepackage{amsmath}

\begin{document}
\bibliographystyle{prsty}
\title{Rare Events Statistics in Reaction--Diffusion Systems}
\draft
\author{Vlad Elgart and Alex Kamenev}

\address{Department of Physics, University of Minnesota,
Minneapolis, MN 55455, USA}

\date{\today}


\begin{abstract}
    {\rm  We develop an efficient method to calculate
    probabilities of large deviations from the typical behavior (rare events) in
    reaction--diffusion systems. The method is based on a semiclassical treatment
    of  underlying "quantum" Hamiltonian, encoding the system's
    evolution. To this end we formulate  corresponding canonical
    dynamical system and investigate its phase portrait. The
    method is presented for a number of pedagogical examples.
      }
\end{abstract}


\maketitle



\bigskip


\section{Introduction}
Reaction--diffusion models have a vast area of applications
\cite{Kampen,Gardiner} ranging from kinetics of chemical reactions
\cite{ZGB86}, biological populations \cite{Murray, Albano94a, Berry}
and epidemics \cite{Liggett85, Mollison77} to the dynamics of financial
markets \cite{BouchaudGeorges90} and ecology \cite{Okubo}.
The models describe dynamics of a number
of particles whose reactions are specified by a certain set or
rules. The rules have a probabilistic nature and are most
conveniently formulated on a lattice in a $d$--dimensional space.
We shall restrict our attention to a wide subclass of such models,
where the particles execute random walks (diffuse) on the lattice,
while the reactions between them are purely local (on--site). Once
the lattice, reaction rules and initial conditions are specified,
one is interested to find statistical characteristics of the
system's subsequent evolution. This goal may be accomplished with
various degrees of detailing and accuracy.

The most detailed information is contained in the probability
distribution functions (PDF) of every possible microscopic state
of the system. The PDF is a solution of an exponentially large
system of  Master equations, which specify probabilities of
transition between every two microscopic states of the system.
Analytical solution of the Master equations is usually unrealistic
and besides the information contained in them is  excessive.
Therefore various approximation schemes are in order. The simplest
one is the mean--field approximation, where a closed set of
equations for average quantities (e.g. concentrations)  is
obtained by an approximate decoupling of higher moments. The
mean--field theory describes a typical evolution of the system, if
the fluctuations are weak in a certain sense. Probability of small
deviations from the mean--field predictions may be found with the
help of Fokker--Planck (FP) equation. It substitutes the discreet
Master equation by a continuum (biased) diffusion equation in the
space of concentrations. Analysis of FP equation is usually
complicated \cite{Kampen,Gardiner}, moreover the approximation is
reliable for small deviations only and fails to provide
probability of large deviations from the typical evolution.

Much attention was attracted recently to reaction--diffusion
systems that are in a close proximity to dynamic
phase--transitions
\cite{OvchinnikovZeldovich78,ToussaintWilczek83,CardyGrassberger85}
(for recent reviews see e.g.
Refs.~[\onlinecite{Hinrichsen,MarroDickman}]). By fine--tuning one
of the parameters some systems may be brought to a point of
quantitative change of their behavior (e.g. stable finite
concentration vs. extinction). In a vicinity of the transition,
neither mean--field nor FP can accurately predict the long--time
scaling of the system's characteristics, such as e.g. particles
concentration. The field--theoretical renormalization group (RG)
methods were developed and successfully applied to a number of
examples \cite{Lee94,CardyTauber96,WOH98,Janssen01}. In
particular, the directed percolation universality class was
identified and studied as the most robust universality class for
the dynamic phase transitions
\cite{Janssen81,CardySugar80,Grassberger97,Hinrichsen}.

In the present work we address somewhat different set of
questions. We consider  a generic reaction--diffusion system that
either does not exhibit, or is far enough from the  phase
transition. A typical evolution scenario and probability of small
deviations are well described by the mean--field theory and the FP
equation. We shall look, however, for a probability of {\em large}
deviations from the typical behavior. "Large" deviation may be
loosely characterized as being of the same order (or larger) as
the typical value (as opposed to the "small" one, which is of the
order of the square root of the typical value). Since the
occurrence of such large deviations has a very small probability,
they my be dubbed as "rare events". Despite being rare the "rare
events" may be of great interest, especially if they cause extreme
consequences. Some of the examples include: proliferation of virus
after immunization (causing death of a patient); large
fluctuations of number of neutrons in a nuclear reactor (causing
explosion), etc. Clearly in these and many other examples one is
interested to know rather precisely how improbable are improbable
events.

Here we develop a rigorous, simple and efficient method to
calculate the rare events statistics in reaction--diffusion
systems. To this end we develop a Hamiltonian formulation of
reaction--diffusion dynamics. Although the system is specified by
a set of rules,  rather than a Hamiltonian, one may nevertheless
show that there is a certain canonical Hamiltonian associated with
the system's dynamics. More precisely, the Master equation may be
reformulated as "quantum" (many--body) Schr\"odinger equation with
some "quantum" Hamiltonian. This observation is not new and is
sometimes referred to as Doi's operator technique
\cite{Doi76,Peliti84}. In fact, its "quantum" version is the basis
for the field--theoretical RG treatment of the dynamical phase
transitions \cite{Lee94,CardyTauber96,WOH98,Janssen01}. Here we
notice that the {\em classical} (or rather semiclassical) dynamics
of the very same Hamiltonian carries a lot of useful information
about reaction--diffusion systems.  In particular, it provides all
the information about the rare events statistics. To extract this
information, it is convenient to formulate the underlying
Hamiltonian in  classical terms (as function of momenta and
coordinates), rather than creation and annihilation operators, as
is customary in the "quantum" approach
\cite{Doi76,Peliti84,MattisGlasser98}.

A particularly convenient tool to visualize the system's dynamics
is a phase portrait of the corresponding Hamiltonian. It consists
of lines (or surfaces) of constant "energy" (the integral of
motion naturally existing in a Hamiltonian system) in the space of
canonical momenta and coordinates. The mean--field (typical)
evolution corresponds to a particular manifold of  zero energy,
given by fixed value of the canonical momenta, $p=1$. Rare events
may be specified by certain initial and finite conditions in the
phase space of the dynamical system, which, in general, do {\em
not} belong to the mean--field manifold. Probability of the rare
event is proportional to $\exp\{-S\}$, where $S$ is  the classical
action on a unique  trajectory, satisfying the specified boundary
conditions. The problem is therefore reduced to finding an
evolution of the {\em classical} dynamical system, whose quantized
Hamiltonian encodes the Master equation. Such task is
substantially simpler  than solving the full "quantum" Master
equation. In fact, even probability of small deviations is much
more efficiently calculated in our semiclassical method, than via
solution of the FP equation (though the latter is also
applicable). For large deviations, however, the FP approach leads
to inaccurate results, while the semiclassical method provides the
simple and accurate prescription. The very similar strategy was
recently applied for the calculation of the full current
statistics of  mesoscopic conductors \cite{Jordan,Bagrets,footJ}.

In this paper we develop the semiclassical method using a number
of reaction--diffusion models as examples. We tried to keep the
presentation self-contained and pedagogical. We start in section
\ref{sectionbinary} from the  model of binary annihilation in zero
dimensions. In section \ref{s3} we complicate the model by
including branching and discuss extinction probability of a system
having a stable population in the mean--field approximation.
Section \ref{secdiffusion} is devoted to the extension of the
formalism to a d--dimensional space. As an example we find an
extinction probability of a finite cluster. In section
\ref{runaway} a population dynamics model with three reaction
channels: reproduction, death and emigration is considered in a
d--dimensional space. The model possesses a long lasting
meta-stable state with a fixed population, that eventually escapes
into the state of unlimited population growth. We show how the
semiclassical method may be used to calculate the lifetime of such
meta-stable state. Finally some conclusions and some open problems
are discussed in section \ref{conclusion}.

\section{Binary Annihilation}
\label{sectionbinary}

The simplest reaction, which we use to introduce notations and set
the stage for  farther discussions, is the binary annihilation
process. It describes a chemical reaction, where two identical
particles, $A$, form a stable aggregate with the probability
$\lambda$ whuch does not involve in further reactions:
$A+A\stackrel{\lambda}{\longrightarrow} \emptyset$. We start from
the zero--dimensional version of the model, where every particle
may react with every other. Such reaction is fully described by
the following Master equation:
\begin{equation}
{d\over dt}P_n(t) = \frac{\lambda}{2}\left[(n + 2)(n + 1)P_{n +
2}(t) - n(n - 1)P_n(t)\right]  ,
                                             \label{master}
\end{equation}
where $P_n(t)$ is a probability to find $n$ particles at time $t$.
The Master equation is to be supplemented with an initial
distribution, e.g. $P_n(0) = e^{-n_0}{n_0^n}/{n!}$ -- the Poisson
distribution  with the mean value $n_0$, or
$P_n(0)=\delta_{n,n_0}$ -- the fixed initial particle number. Let
us define now the {\em generating function} as
\begin{equation}
                                              \label{generating}
  G(p,t) \equiv \sum\limits_{n=0}^\infty p^n P_n(t)\, .
\end{equation}
Knowing the generating function, one may find a probability of
having (integer) $n$ particles at time $t$ as $P_n(t)=
\partial_p^n G(p,t)|_{p=0}/n!$. If $n\gg 1$ it is more convenient,
to use an alternative representation:
\begin{equation}
                                                \label{finiteprob}
P_{n}(t) = {1\over 2\pi i} \oint {dp\over p}\,
G(p,t)\, p^{-n}\, ,
\end{equation}
where the integration is performed over a closed contour on the
complex $p$~--~plane, encircling $p=0$ and going through the
region of analyticity of $G(p,t)$.

The point $p=1$ plays a special role in this formulation. First of
all, the conservation of probability demands the fundamental
normalization condition:
\begin{equation}
G(1,t)\equiv 1\, .
                                                     \label{norm}
\end{equation}
Second, the moments of the PDF, $P_n(t)$, may be expressed through
derivatives of the generating function at $p=1$, e.g. $\langle
n(t)\rangle \equiv\sum\limits_n nP_n(t) =\partial_p
G(p,t)|_{p=1}$.

In terms of the generating function the Master
equation~(\ref{master}) may be {\em identically} rewritten as
\begin{equation}
{\partial G\over \partial t} = -{\lambda\over 2}(p^2 -1)\,
{\partial^2 G \over \partial p^2}\, .
                                      \label{genmaster}
\end{equation}
This equation is to be solved with some initial condition, e.g.
$G(p,0)=\exp\{n_0(p-1)\}$ for the Poisson initial distribution or
$G(p,0)=p^{n_0}$ for rigidly fixed initial particle number.  The
solution should  satisfy the normalization condition,
Eq~(\ref{norm}), at any time. In addition, all physically
acceptable solutions must have all $p$--derivatives at $p=0$
non-negative.

One may consider Eq.~(\ref{genmaster}) as the "Schr\"odinger"
equation:
\begin{equation}
{\partial\over \partial t}\, G = -\hat H G\, ,
                             \label{shrodinger}
\end{equation}
where the "quantum" Hamiltonian operator, $\hat H$, in the $\hat
p$ ("momentum") representation is:
\begin{equation}
\hat H(\hat p,\hat q) = {\lambda\over 2}(\hat p^2 -1)\hat q^2\, .
                                       \label{hamiltonian}
\end{equation}
Here we have introduced the "coordinate" operator, $\hat q$ as
\begin{equation}
\hat q\equiv - {\partial\over \partial p}\, ; \hskip 2cm [\hat
p,\hat q]=1\,.
                                          \label{commutator}
\end{equation}
The "Hamiltonian", Eq.~(\ref{hamiltonian}), is normally ordered
and not Hermitian. The last fact does not present any significant
difficulties, however.

If the "quantum" fluctuations are weak (which in present case is
true as long as $\langle n(t)\rangle\gg 1$), one may employ the
WKB approximation to solve the "Schr\"odinger"--Master equation.
Using anzatz $G(p,t)=\exp\{-S(p,t)\}$ and expanding $S(p,t)$ to
the leading order in $1/\lambda$, one obtains the classical
Hamilton--Jacoby equation:
\begin{equation}
{\partial S\over \partial t} = H\left(p,{\partial S \over \partial
p} \right)={\lambda\over 2}(p^2 -1)\, \left({\partial S \over
\partial p}\right)^2 \, .
                                          \label{hamiltonjacoby}
\end{equation}
Instead of directly solving the Hamilton-Jacoby  equation we'll
develop the  Hamilton approach, which is much more convenient for
finite dimensional applications.

To this end we employ the Feynman path integral representation,
which may be derived  introducing resolution of unity at each
infinitesimal time--step and employing the normal ordering. As a
result, one finds for the generating function:
\begin{equation}
G(p,t) = \lim\limits_{M\to \infty}\int \!\prod\limits_{k=0}^{M}
{dp_k d q_k\over 2\pi}\,
  e^{-S[p_k,q_k]}\,,
                                          \label{pathintegral}
\end{equation}
where  the discrete representation for the action $S[p_k,q_k]$ is
given by
\begin{eqnarray}
  S = &&\sum\limits_{k=1}^{M}\left[ p_k(q_k - q_{k-1}) + H(p_{k},q_{k-1})\delta
t\right]
  \nonumber \\
  &&+\, p_0 q_0  - p q_M  - n_0(p_0-1)\,
                                           \label{action}
\end{eqnarray}
and $\delta t = t/(M+1)$. The last term in this expression is
specific to the Poisson initial conditions. If the initial number
of particles is fixed to be $n_0$, and therefore $G(p,0)=p^{n_0}$
-- the last term is changed to $n_0\ln p_0$.  The same path
integral may be derived, of course, using the Doi's operator
algebra and coherent states. We summarize this derivation in
Appendix \ref{appendix}. The convergency of the path integral may
be achieved by a proper rotation in the complex $p_k$ and $q_k$
planes.

In what follows we are interested in the semiclassical treatment
of this path integral. Varying  the action with respect to $p_k$
and $q_k$ for $k=0,1\ldots M$, one obtains the classical equations
of motion (in continuous notations):
\begin{subequations}
\label{dot}
\begin{eqnarray}
  \dot{q} &=& -{\partial H\over \partial p} = -\lambda p\, q^2\, ; \label{dotq}
\\
  \dot{p} &=& {\partial H\over \partial q} =\lambda (p^2-1) q\,   \label{dotp}
\end{eqnarray}
\end{subequations}
and the boundary conditions:
 \bs
 \label{boundary}
\begin{eqnarray}
  q(0) &=& n_0\, ; \label{q0} \\
  p(t)&=& p\, , \label{pT}
\end{eqnarray}
 \es
where $p$ and $t$ are the arguments of the generating function
$G(p,t)$. Notice that while the coordinate is fixed at an initial
time (past), the momentum is imposed at a finite time (future).
These equations admit the integral of motion, which we call
"energy": $\dot E=0$, where
\begin{equation}
                                              \label{3}
  E\equiv H(p(t),q(t)) = \frac{\lambda}{2} (p^2(t)-1) q^2(t)\, .
\end{equation}
As a result, the action {\em on a classical trajectory} may be
written as (in continuous notations):
\begin{equation}
  S[p,q] = E t  -\int\limits_0^t\!\! q \dot{p}\,\mathrm{d}t  - n_0(p(0)-1)\, .
                                              \label{classaction}
\end{equation}

\begin{figure}
\includegraphics[width=8cm]{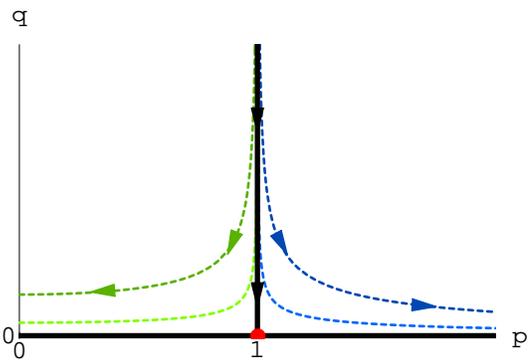}
\caption{The phase portrait of the binary annihilation process.
Thick lines represent solution of $H(p,q) = 0$, fat dot -- fixed
point. Thinner lines represent dynamical trajectories with
non--zero energy. Line $p=1$ gives the mean--field dynamics.}
\label{bin}
\end{figure}

To find the low moments one needs to know $G(p,t)$ in the
immediate vicinity of $p=1$. In this  case the Hamilton
equations~(\ref{dot}) with the boundary
conditions~(\ref{boundary}) may be solved with the {\em
mean--field} anzatz:
 \bs
 \label{meanfield}
\begin{eqnarray}
&&\bar p(t)\equiv 1\, ;\label{meanfieldp}   \\
&&{d \bar q\over dt} = -\left. {\partial H\over \partial
p}\right|_{p=1} = -\lambda \bar q^2\, .
                           \label{meanfieldq}
\end{eqnarray}
 \es
The last equation constitutes the mean--field approximation for
the reaction coordinate, $\bar q\equiv \bar n\approx \langle
n\rangle$. The classical action, Eq.~(\ref{classaction}), is
obviously nullified on the mean--field solution: $S[\bar p,\bar
q]=0$. This enforces the normalization, Eq.~(\ref{norm}) (it is
straightforward to show that the fluctuation determinant around
the mean--field trajectory is unity). In fact, any legitimate
Hamiltonian must satisfy the condition $H(1,q)=0$ to insure
normalization. As a result, the mean--field solution, $p=1$, is
bound to have zero energy, $\bar E\equiv 0$.

However, the assumption that $p=1$ is {\em not} always a
legitimate one. The probability of any event other than  the
mean-field prediction is automatically described by $p(t)=p$
different from unity. The rare events definitely belong to this
category. For such cases the mean--field anzatz,
Eq.~(\ref{meanfield}), is {\em not} applicable and one must go
back to the full dynamical system, Eqs.~(\ref{dot}) (provided
semiclassical approximation is justified). For example, let us
imagine doing the contour integral, Eq.~(\ref{finiteprob}), by the
stationary point method. Approximating, $G(p,t)=\exp\{-S(p,t)\}$,
with the classical action, $S$, one finds for the saddle point
condition: $n = -p\,\partial S/\partial p = p(t)q(t)$, where we
have used that on a classical trajectory $\partial S/\partial p =
- q(t)$. Therefore, if one is interested in $n$ which is different
from the mean--field prediction $\bar q(t)$, one must consider
$p(t)=p$ to be different from unity.

In case of the binary annihilation the mean--field prediction is
$\bar q(t)\equiv \bar n(t) = n_0/(1+n_0\lambda t)\approx (\lambda
t)^{-1}$ for $1<(\lambda t)^{-1}\ll n_0$. We are looking for a
probability to find $n\neq \bar n(t) =(\lambda t)^{-1}$ particles
at time $t\gg (\lambda n_0)^{-1}$. The phase portrait of the
dynamical system, Eqs.~(\ref{dot}), is plotted on Fig.~\ref{bin}.
Dynamical trajectories for a given energy, $E$, are given by
$q=\sqrt{2 E\lambda^{-1}/(p^2-1)}$. Since $q(0) = n_0\gg 1$, one
finds $p(0)= 1+2E/(\lambda n_0^2)\approx 1$. Substituting this
trajectory into Eq.~(\ref{dotp}) and integrating it between
$p(0)\approx 1$ and $p(t)=p$, one finds $ E = -\arccos^2
p/(2\lambda t^2)$. The corresponding classical action,
Eq.~(\ref{classaction}), is given by
\begin{equation}
  S(p,t) = {1\over 2}\, \bar n(t) \arccos^2 p  \, .
                                    \label{binoryaction}
\end{equation}
This action solves the Hamilton--Jacoby equation
(\ref{hamiltonjacoby}) and is nullified at the mean--field
trajectory, $p=1$. As a result, the generating function is given
by $G(p,t)\approx \exp\{-S(p,t)\}$ with the classical action,
Eq.~(\ref{binoryaction}).

We are now on the position to find the rare events statistics:
namely we are looking for the probability to find $n$ particles
after time $t$, that is $P_n(t)$, where $n$ is significantly
different from the mean field prediction $\bar n=(\lambda
t)^{-1}$. To this end one may perform integration, required by
Eq.~(\ref{finiteprob}), in the stationary point approximation to
obtain for the probability distribution
\begin{equation}
P_n(t)={\cal N} \exp\left\{-\bar n\left({1\over 2} \arccos^2 p_s
+{n\over \bar n} \ln p_s\right)\right\}\, ,
                                                       \label{binorudistr}
\end{equation}
where $p_s=p_s(n/\bar n)$ is the solution of the saddle point
equation: $p_s(p_s^2-1)^{-1/2}\arccos p_s = n/\bar n$. In the
limiting cases the  exponent takes the form:
\begin{equation}
-\ln P_n(t) \approx \left\{ \begin{array}{ll} {\pi^2\over 8}\bar n
-n\ln {\pi\bar n\over 2n}  \,; & n\ll \bar n\, ,\\
 {3\over
4}\,(n-\bar n)^2/\bar n \,; \,\,\,\,& |n-\bar n|\ll \bar n\, , \\
{1\over 2}n^2/\bar n-n\ln 2 \,; & n\gg \bar n  \, .
 \end{array} \right.
                                \label{binarytails}
\end{equation}
\begin{figure}
\includegraphics[width=8cm]{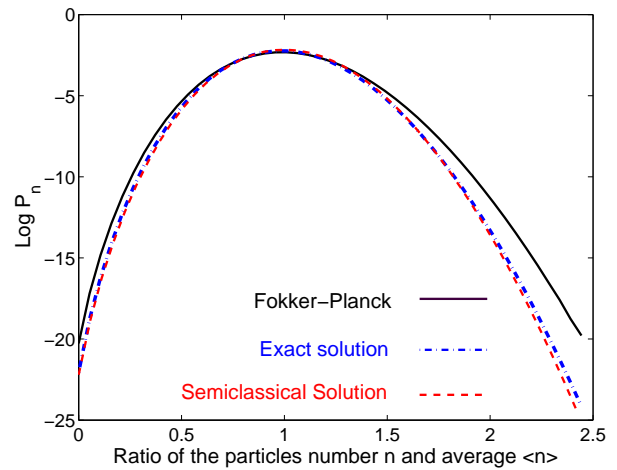}
\caption{Logarithm of the PDF $P_n(t)$ as function of $n/\bar n
(t)$ at a fixed $t$. The semiclassical result,
Eq.~(\ref{binorudistr}), -- dashed line; the exact solution
produced by the numerical simulation of the Master equation
(\ref{master}) -- dashed-dotted line; numerical solution of the FP
equation: $\dot P = \lambda[ ((n^2-n)P)'' + (n^2P)']$  -- full
line.}\label{fp}
\end{figure}
The logarithm of the PDF  is plotted on Fig. \ref{fp} versus
$n/\bar n$ for a fixed $\bar n=\bar n(t)$. The corresponding
exponent resulting from the solution of the Focker--Planck
equation is shown on the same plot for comparison. The two
exponents coincide for small deviations from the mean--field
result, $|n/\bar n -1|\ll 1$. For larger deviations (rare events),
$n/\bar n \sim O(1)$, the Focker--Planck results are significantly
off the correct ones. Finally, the normalization factor ${\cal
N}=\sqrt{3/(4\pi \bar n)}$ is simply determined by the immediate
vicinity of the maximum of the distribution, $|n-\bar n|\ll \bar
n$.

\section{Branching and Annihilation}
\label{s3}

Let us consider now a more interesting example of binary
annihilation with branching. The model consists of the two
reactions: annihilation $A+A\stackrel{\lambda}{\longrightarrow}
\emptyset$ and branching $A\stackrel{\sigma}{\longrightarrow} 2A$.
The Master equation is written as:
\begin{eqnarray}
{d\over dt}P_n(t) &=& \frac{\lambda}{2}\left[(n + 2)(n + 1)P_{n +
2}(t) - n(n - 1)P_n(t)\right]  \nonumber \\
 &+& \sigma\left[(n - 1)P_{n -
1}(t) - nP_n(t)\right]\,  ,
                                             \label{masterbranching}
\end{eqnarray}
one may  check that the  corresponding Hamiltonian takes the form:
\begin{equation}
  \hat H(\hat p,\hat q) = \frac{\lambda}{2} (\hat p^2 - 1) \hat q^2 -
  \sigma(\hat p - 1) \hat p \hat q\, .
                                             \label{H-branching}
\end{equation}
As expected, it satisfies the normalization condition, $H(1,q)=0$.
The classical equations of motion are
 \bs
 \label{dotb}
\begin{eqnarray}
  \dot{q} &=&  -\lambda p\, q^2+\sigma(2p-1)q\, ; \label{dotqb}  \\
  \dot{p} &=& \lambda (p^2-1) q-\sigma(p-1)p\, ,  \label{dotpb}
\end{eqnarray}
 \es
with the same  boundary conditions as in the previous example,
Eqs.~(\ref{boundary}). The classically conserved energy is
$E=H(p(t),q(t))$. The mean--field anzatz, $\bar p(t)\equiv 1$,
leads to the  mean--field equation for the reaction coordinate,
$\bar q\approx \langle n\rangle$:
\begin{equation}
{d\bar q\over dt} =  -\lambda \bar q^2+\sigma \bar q\, .
                                                   \label{meanfieldb}
\end{equation}
This equation possesses  two stationary states: the active one
$\bar q=\sigma/\lambda\equiv n_s$ and the passive one $\bar q=0$.
Below we show that the active state is {\em not} actually
thermodynamically stable (in $0d$ system) and in a finite time
decays into the passive one.

\begin{figure}
\includegraphics[width=8cm]{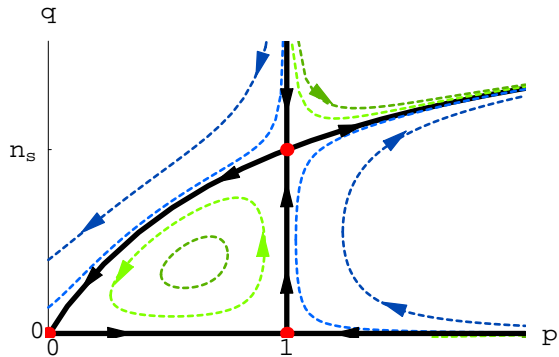}
\caption{The phase portrait of the branching annihilation process.
Thick  lines are lines of zero energy, $H(p,q) = 0$. Fat dots are
fixed points.}\label{creat_ann}
\end{figure}

To proceed with the discussion of the rare events statistics, we
need a phase portrait of the system. It contains three lines of
zero energy: the mean--field one $p=1$; the empty system one $q=0$
and the non--trivial line $q =2 n_s p/(1 + p)$. These lines
determine the topology of the phase diagram, Fig.~\ref{creat_ann},
where the arrows show the positive time direction. According to
the mean--field equation (\ref{meanfieldb}),  from any initial
state with $n_0\neq 0$, the system reaches the active state with
$n_s$ particles during the time $t\approx \sigma^{-1}$. Hereafter
we assume that $n_s=\sigma/\lambda\gg 1$. We shall look for a
probability to find $n\neq n_s$ particles after a time $t\gg
\sigma^{-1}$.

Of particular interest, of course, is the probability of going to
the  passive state, namely $n=0$, during a large time $t$.
According to the definition of the generating function,
Eq.~(\ref{generating}), this probability is given by $G(0,t)$. We
are interested, therefore, in the trajectory which starts at some
initial coordinate $q_0=n_0$ (and arbitrary momentum) and ends at
$p_M=0$ (and arbitrary coordinate) after time $t$. In a long time
limit, $t\to \infty$, such trajectory approach the lines of zero
energy. The system first evolves along the mean--field trajectory
$p=1$ towards the active state, $q=n_s$, and then goes along the
non--trivial line $q =2 n_s p/(1 + p)$ towards the passive state
$p=q=0$, cf. Fig.~\ref{creat_ann}. The action is zero on the
mean--field part of the evolution, while it is
\begin{equation}
S_0=-\int\limits_1^0 \frac{2n_s p}{1+p}\, dp = n_s2(1-\ln 2)
                                               \label{decayaction}
\end{equation}
along the non--trivial line.

According to the standard semiclassical description of tunnelling
\cite{Zinn-Justin}, to find an escape probability, one
has to sum up contributions of all classical trajectories with an
arbitrary number of bounces from $(1,n_s)$ to $(0,0)$ and back.
Each bounce brings the factor $\sigma t e^{-S_0}$, where the
pre-factor reflects the fact the center of the bounce may take
place at any time without changing the action (zero mode). Since
the distant (in time) bounces interact with each other only
exponentially weak, the escape attempts are practically
uncorrelated. As a result, the probability to find an empty
system, $P_0(t) =G(0,t)$, is
\begin{equation}
P_0(t)  = 1-e^{-t/\tau}\, ,
                                                \label{decay}
\end{equation}
where the decay time $\tau$ is given by
\begin{equation}
\tau = \sigma^{-1} \exp\{+S_0\}\, .
                                                 \label{decaytime}
\end{equation}
The semiclassical calculations is valid as long as $S_0\gg 1$ and
thus the decay time is much longer than the microscopic time,
$\tau\gg \sigma^{-1}$.


\section{Diffusion}
\label{secdiffusion}

We turn now to the discussion of finite dimensional systems. To
characterize a microscopic state one need to specify number of
particles at every site of the lattice: $\{n_1,\ldots, n_N\}$,
where $N\sim L^d$ is the total number of sites. The probability of
a given microscopic state may be written as $P_{n_1,\ldots,
n_N}(t)$ and the corresponding generating function is
\begin{equation}
G(p_1,\ldots, p_N,t)\equiv \sum\limits_{n_1,\ldots,n_N}
p_1^{n_1}\ldots p_N^{n_N}P_{n_1,\ldots, n_N}(t)\,.
                                      \label{genear-diff}
\end{equation}
Assuming that the reaction rules are purely local (on-site), while
the motion on the lattice is diffusive, one finds that the
Hamiltonian takes the form
\begin{equation}
\hat H(\hat p_1,\ldots, \hat p_N,\hat q_1,\ldots,\hat
q_N)\!=\!\sum\limits_i\left[ \hat H_0(\hat p_i,\hat q_i) + D\nabla
\hat p_i\!\cdot\!\nabla \hat q_i\right],
                                          \label{H-diff}
\end{equation}
where $\hat H_0(\hat p,\hat q)$ is a zero--dimensional on--site
Hamiltonian given e.g. by Eqs.~(\ref{hamiltonian}) or
(\ref{H-branching}); $D$ is a diffusion constant and $\nabla$ is
the lattice gradient. To shorten notations we  pass to  continuous
$d$--dimensional variable $x$ and introduce fields $p(x)$ and
$q(x)$. The generating function becomes generating functional,
$G(p(x),t)$. The latter may be  written as a functional integral
over canonically conjugated  fields $p(x,t)$ and $q(x,t)$, living
in $d+1$ dimensional space, with the action
\begin{equation}
  S[p,q] = \int\limits_0^t\!\! dt\! \int \!\! d^dx \left[H_0(p,q)+D\nabla p\cdot
\nabla q-
  q \dot{p}\,\right] \, .
                                              \label{classaction-diff}
\end{equation}
The initial term, e.g. the Poisson one:  $\int d^d x\,
n_0(x)(1-p(x,0))$, should also be added to the action. The
corresponding classical equations of motions are:
 \bs
 \label{dot-diff}
\begin{eqnarray}
  \dot{q} &=& D\nabla^2 q -{\delta H_0\over \delta p} \, ; \label{dotq-diff} \\
  \dot{p} &=& -D\nabla^2 p + {\delta H_0\over \delta q}\, .    \label{dotp-diff}
\end{eqnarray}
 \es
These equations are to be solved with the following boundary
conditions:
 \bs
 \label{boundaryx}
\begin{eqnarray}
  q(x,0) &=& n_0(x)\, ; \label{q0-diff} \\
  p(x,t)&=& p(x)\, , \label{pT-diff}
\end{eqnarray}
 \es
where $n_0(x)$ is an initial space--dependent concentration and
$p(x)$ is the source field in the generating functional
$G(p(x),t)$. The mean--field approximation is obtained by putting
$p(x,t)=1$ and is described by the reaction--diffusion equation:
\begin{equation}
\partial_t\bar q=D\nabla^2 \bar q-
\left.\frac{\delta H_0(p,\bar q)}{\delta p}\right|_{p=\bar p=1}\,
,
                                                   \label{reactiondiffusion}
\end{equation}
that is subject of numerous studies.

Equations (\ref{dot-diff})  admit the integral of motion:
$E=\int\! d^d x\, [H_0(p,q)+D\nabla p \nabla q]$. In some cases
(see below) an additional infinite sequence of integrals of motion
may be found, making the classical problem, Eqs.~(\ref{dot-diff}),
analytically solvable. In a general case, these equations must be
solved numerically. We notice, however, that such numerical
problem is orders of magnitude simpler than numerical solution of
the Master and even FP equations, or direct modelling of the
stochastic system. Below we discuss a fast, efficient algorithm
for numerical solution of Eqs.~(\ref{dot-diff})  with the boundary
conditions Eqs.~(\ref{boundaryx}). Moreover, a lot of insight may
be gained by investigating the phase portrait of the
zero--dimensional Hamiltonian, $H_0(p,q)$, which allows to make
some semi--quantitative predictions without numerical solution.

To illustrate how the method works we consider the branching
annihilation problem of section \ref{s3} ($H_0$ is given by
Eq.~(\ref{H-branching})) on a compact $d$--dimensional cluster --
the "refuge" \cite{Lindenberg03}, denoted as ${\cal R}$. Outside
of the refuge, there is a very high mortality, $A\to \emptyset$,
rate which is eventually taken to infinity. This dictates the
boundary condition
\begin{equation}
q(\partial{\cal R},t)=0\, ,
                     \label{refugebound}
\end{equation}
where $\partial{\cal R}$ is the boundary of the cluster ${\cal
R}$. It is convenient to pass to the dimensionless time $\sigma
t\to t$ and coordinates $x/\xi\to x$, where $\xi\equiv
\sqrt{D/\sigma}$. We also introduce the rescaled fields
$q(x,t)=n_s\varphi(x,t)$ (where $n_s=\sigma/\lambda$) and
$p(x,t)=1-\hat \varphi(x,t)$. In these notations the semiclassical
equations, Eq.~(\ref{dot-diff}), take the symmetric form:
 \bs
 \label{vu}
\begin{eqnarray}
  \partial_t \varphi = \nabla^2 \varphi + \varphi -  \varphi^2 +
  \hat\varphi  \varphi^2 - 2  \hat\varphi  \varphi\,,\label{v}\\
  -\partial_t  \hat\varphi = \nabla^2  \hat\varphi +  \hat\varphi -
  \hat\varphi^2 +  \varphi  \hat\varphi^2 - 2  \varphi  \hat\varphi\,,\label{u}
\end{eqnarray}
 \es

Consider first the mean--field ($ \hat\varphi=0$) evolution,
described by the equation
\begin{equation}
  \partial_t  \varphi=  \nabla^2  \varphi +  \varphi -  \varphi^2\, .
                                   \label{meanfieldrefuge}
\end{equation}
subject to the boundary condition $ \varphi(\partial{\cal
R},t)=0$. For the small concentrations, $ \varphi\ll 1$, the last
term may be omitted and the solution takes the form:
\begin{equation}
                             \label{refugeY}
   \varphi(x,t) = \sum\limits_{n=0}^\infty \alpha_n
   e^{\,(1-\lambda_n) t}\, Y_n(x)\,,
\end{equation}
where $Y_n(x)$ are normalized eigenfunctions of the Laplace
operator in the region $\,{\cal R}\,$ with zero boundary
conditions and eigenvalues $-\lambda_n < 0$;  coefficients
$\alpha_n$ depend on an initial condition. Therefore, if the
smallest eigenvalue, $\lambda_0$, is larger than unity (the
cluster is small enough), any initial distribution evolves towards
the empty system. The characteristic lifetime of the system is
thus
\begin{equation}
\tau=\sigma^{-1}(\lambda_0-1)^{-1}\,; \hskip 1cm  \lambda_0>1\, .
                                  \label{smallrefugelife}
\end{equation}

If $\lambda_0<1$ (the cluster is larger than some critical size),
the mean--field evolution, Eq.~(\ref{meanfieldrefuge}), leads to a
stable non--vanishing concentration $ \varphi_0(r)$, which is
given by  the solution of the equation $\nabla^2  \varphi_0 +
\varphi_0 -  \varphi_0^2=0$ with zero boundary conditions. It is
clear, however, that such solution is actually a meta-stable state
of the system. Namely, after a long enough time the system will
find itself in the empty (passive)  state. Our task is to find the
system's lifetime, $\tau$, for the meta-stable case,
$\lambda_0<1$. According to our previous discussions the lifetime
is expected to be exponentially long
\begin{equation}
\tau = \sigma^{-1} e^{\,S_d} \,; \hskip 1cm  \lambda_0<1\, ,
                                  \label{largerefugelife}
\end{equation}
where $S_d$ is the action along the semiclassical trajectory, that
solves Eqs.~(\ref{v}), (\ref{u}) with the initial condition $
\varphi(x,0)= \varphi_0(x)$ and the final condition
$\hat\varphi(x,t_e)=1$. The extinction time, $t_e$, is to be sent
to infinity. Indeed $\partial S_d/\partial t_e=E(t_e)\leq 0$ and
thus the longer the extinction time -- the smaller the action. In
practice, however, the action almost saturates at modest values of
$t_e$.

In general the problem cannot be solved analytically and one needs
to resort to numerical approaches. The following iteration scheme
rapidly converges to the desired solution: one first fixes momenta
to be $\hat\varphi_1(x,t)=1$ at any time and solves Eq.~(\ref{v})
with the initial condition $ \varphi(x,0)= \varphi_0(x)$ by
forward iteration from $t=0$ to $t=t_e$. The result of this
procedure, $\varphi_1(x,t)$, is kept fixed during the next step,
that is solution of Eq.~(\ref{u}) with the condition $\hat
\varphi(x,t_e)=1$ by the backward iteration from $t=t_e$ to $t=0$.
This way one finds $\hat\varphi_2(x,t)$, which is kept fixed while
the next approximation $\varphi_2(x,t)$ is obtained by the forward
iteration of Eq.~(\ref{v}). Repeating successively  forward and
backward iterations the algorithm rapidly converges to the
required solution. The action $S_d=S_d(t_e)$ is then calculated
according to Eq.~(\ref{classaction-diff}). Finally one has to
check that $S_d(t_e)$ does not decrease significantly upon
increasing $t_e$.

\begin{figure}
\includegraphics[width=8cm]{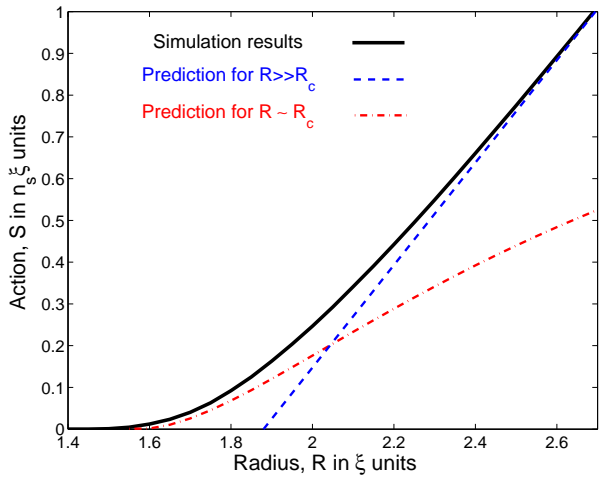}
\caption{The semiclassical action (in units of $\xi^d n_s$) for
the extinction of a one dimensional cluster is shown as function
of cluster's radius $R$ (in units of $\xi$) -- full line. Large
radius approximation, Eq.~(\ref{largeclusteraction}), is shown by
the dashed line; near--critical, $\pi/2\lesssim R$, approximation,
Eq.~(\ref{smallclusteraction}) -- dashed-dotted line. }
\label{fig-refuge}
\end{figure}

The action, $S_d$, for  one--dimensional cluster of size $2R$ is
plotted on Fig.~\ref{fig-refuge} as function of $R$. The critical
radius $R_c=\pi/2$ (in units of $\xi$) is found from the condition
$\lambda_0=1$. For $R<R_c$, there is no meta-stable state and thus
$S_d=0$, while the cluster lifetime is given by
Eq.~(\ref{smallrefugelife}). For $R>R_c$ the lifetime is given by
Eq.~(\ref{largerefugelife}), with the numerically calculated $S_d$
plotted on Fig.~\ref{fig-refuge}. The asymptotic behavior of the
action $S_d$ for $R\gg R_c$ ($\lambda_0\ll 1$) and $R-R_c\ll R_c$
($1-\lambda_0\ll 1$) may be readily found analytically.

For $R\gg R_c$  ($\lambda_0\ll 1$) the concentration throughout
the  bulk of the cluster is practically uniform apart from a
surface layer of thickness $\xi$. One may therefore apply the
results of the zero dimensional problem, Eq.~(\ref{decayaction}),
to find
\begin{equation}
S_d= 2(1-\ln 2)n_s\xi^d\left({\cal V} - c{\cal S}\right)\, ,
                                    \label{largeclusteraction}
\end{equation}
where ${\cal V}$ and ${\cal S}$ are cluster's dimensionless volume
and surface area correspondingly and $c$ is  a numerical constant,
which we shall not  evaluate here. For the 1d case the
corresponding line is plotted on Fig.~\ref{fig-refuge} by the
dashed line.

We turn finally to the clusters that are only slightly larger than
the critical size: $\epsilon\equiv 1-\lambda_0\ll 1$ (e.g for a
spherical cluster with the radius $R$ one finds $\epsilon =
1-(R_c/R)^2$, where $R_c$ is a critical radius, found from
$\lambda_0=1$). In this case only the zeroth eigenfunction
$Y_0(x)$ is the unstable direction of the linearized mean--field
equation. The full non--linear mean--field
equation~(\ref{meanfieldrefuge}) possesses, therefore, the stable
solution $ \varphi_0(x)$, that is expected to be of order
$\epsilon$. One may thus look for this solution in the following
form:
\begin{equation}
 \varphi_0(r)=\epsilon\left( \eta Y_0(x) + \epsilon  \varphi_1(x)\right) \, ,
                                            \label{critical-cluster}
\end{equation}
where $ \varphi_1(x)$ is orthogonal to $Y_0(x)$. One can now
substitute this trial solution in Eq.~(\ref{meanfieldrefuge}),
keeping only the leading (second) order in $\epsilon$, and project
on $Y_0$, using its orthogonality to $ \varphi_1$. As a result the
coefficient $\eta$ is found to be:
\begin{equation}
\eta^{-1}=\int\limits_{\cal R}\!\! d^dr\, Y_0^3(r)\, .
                                        \label{eta}
\end{equation}
The meta-stable  solution of equations (\ref{vu}) in the leading
order in $\epsilon$ is therefore: $ \varphi_0(x)=\epsilon \eta
Y_0(x)$ and $\hat\varphi_0(x)=0$. To find the optimal escape
trajectory, let us parameterize deviations from this meta-stable
state as
 \bs
 \label{vu-div}
\begin{eqnarray}
 \varphi(x,t) &=&\epsilon \eta Y_0(x) +\sum\limits_{n=0}^\infty
\alpha_n(t)\,Y_n(x)\,; \label{v-div}\\
\hat\varphi(x,t) &=& \sum\limits_{n=0}^\infty \beta_n(t)\,
Y_n(x)\,,
                                                 \label{u-div}
\end{eqnarray}
 \es
where $\alpha_n(t)$ and $\beta_n(t)$ are assumed to be small. One
can now substitute these deviations into the dynamical equations
(\ref{vu})  and linearize them with respect to $\alpha_n,\,\,
\beta_n$. It is straightforward to see then that in the leading
order in $\epsilon$ only $\alpha_0$ and  $\beta_0$ should be
retained. They evolve according to:
\begin{equation}
\frac{d}{d t} \left(
\begin{array}{c}
\alpha_0 \\ \beta_0
\end{array}\right) =\epsilon
\left(\begin{array}{cc} -1  & -2\\ 0& 1
\end{array}\right)
\left(\begin{array}{c} \alpha_0 \\ \beta_0
\end{array}\right) + O(\epsilon^2)\,.
                                             \label{alphabeta}
\end{equation}

\begin{figure}
\includegraphics[width=8cm]{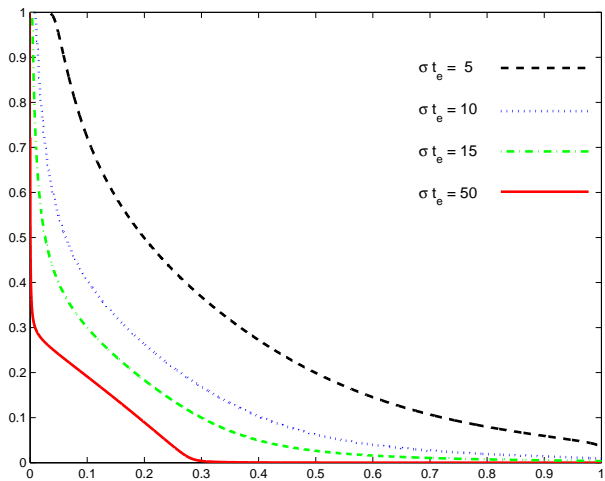}
\caption{Trajectories on the $(\hat\varphi,\varphi)$ plane is
shown for the center point of the 1d cluster, $x=0$. The
horizontal axes is $\hat\varphi$ variable and the vertical axes is
$\varphi$. Different curves are distinguished by their escape
time, marked in dimensionless units $\sigma t_e$. $\epsilon\eta
Y_0(0)=0.3$.} \label{fig-escape}
\end{figure}

The matrix on the right hand side has two eigenvectors, $(1,0)$
and $(1,-1)$  with the eigenvalues $-1$ and $1$ correspondingly.
The first one describes  deviation in the mean-field direction,
$\hat\varphi=0$, and leads to the restoring force back to the
meta-stable state. The second eigenvector  gives the most unstable
direction, that describes the way the system escapes towards the
empty state. The corresponding trajectory on the $(\hat\varphi,
\varphi)$ plane is plotted on Fig.~\ref{fig-escape} for the center
point of the 1d cluster, $x=0$. Different lines correspond to  a
few values of $t_e$. For $t_e\to \infty$ the energy, $E$,
approaches zero and the trajectory approaches the $(1,-1)$
direction that leads from the meta-stable point $(0,\epsilon\eta
Y_0)$ to a symmetric meta-stable point $(\epsilon\eta Y_0,0)$.
(The existence of the latter follows directly from the symmetry of
equations (\ref{vu}).) For $\epsilon\ll 1$ the small deviation
analysis describe the entire transition between the two
meta-stable points, that takes place, therefore, along the
straight line:
\begin{equation}
 \varphi(x,t)=\epsilon\eta\,Y_0(x) - \hat\varphi(x,t)\, ,
                                          \label{escapetrajectory}
\end{equation}
on the $(\hat\varphi,\varphi)$ plane. The farther evolution takes
place along $\varphi=0$ direction. As a result, in the limit
$t_e\to\infty$, and therefore $E\to 0$, the semiclassical escape
action  is given by the area of the straight triangle with the
hight $\epsilon\eta Y_0(x)$ integrated over the cluster, cf.
Eq.~(\ref{classaction-diff}),
\begin{equation}
S_d=-\int\limits_{\cal R}\!\!d^dx\!\!\int\!\! qdp=n_s\xi^d
\int\limits_{\cal R}\!\!d^dx\!\!\int\!\!  \varphi d\hat\varphi =
 {1\over 2} (\eta\epsilon)^2 n_s\xi^d\, .
                                       \label{smallclusteraction}
\end{equation}
For the 1d cluster ($\epsilon=1-(\pi/2R)^2$ and
$\eta^2=9\pi^3/128$) Eq.~(\ref{smallclusteraction}) is shown on
Fig.~\ref{fig-refuge} by a dashed--doted line. (For a circular
cluster in 2d $R_c=2.4$ and $\eta^{2} = 9.4$, while for a
spherical 3d cluster $R_c=\pi$ and $\eta^{2} = 51.7$.) One may
observe that the large and small cluster asymptotic results,
Eqs.~(\ref{largeclusteraction}) and (\ref{smallclusteraction})
correspondingly, provide a reasonable approximation for the exact
numerical calculation of the semiclassical action, $S_d$. Finally,
the probability of the system staying in the meta-stable state is
$P(t)=\exp\{-t/\tau\}$, where the lifetime $\tau$ is given by
Eq.~(\ref{largerefugelife}).

\section{Run--away systems}
\label{runaway}

In this section we consider a qualitatively different system that
exhibits a run--away behavior, characterized by unlimited
proliferation of the number of particles. The simplest example is
given by the population dynamics model consisting of three
reactions: binary reproduction, death and emigration,
characterized by probabilities $\lambda$, $\sigma$ and $\mu$
correspondingly. The schematic way to write it is:
$A+A\stackrel{\lambda}{\longrightarrow} 3A$;
$A\stackrel{\sigma}{\longrightarrow} \emptyset $ and $\emptyset
\stackrel{\mu}{\longrightarrow} A$. The Master equation for the
zero--dimensional system has the form:
\begin{eqnarray}
{d P_n\over dt} &=& \lambda\left[\frac{(n -1)(n -2)}{2}\, P_{n -1}
- \frac{n(n - 1)}{2}\, P_n\right] \nonumber \\
&+&\sigma\left[ (n+1)P_{n+1}-nP_n\right] + \mu
\left[P_{n-1}-P_n\right] .
                                             \label{master-run}
\end{eqnarray}
The corresponding zero--dimensional Hamiltonian is:
\begin{equation}
\hat H_0(\hat p,\hat q)={\lambda\over 2}(\hat p^2-\hat p^3)\hat
q^2+ \sigma(\hat p-1)\hat q+\mu(1-\hat p)\,
                                             \label{H-run}
\end{equation}
and the classical equations of motions are:
 \bs
\label{dot-run}
\begin{eqnarray}
  \dot{q} &=&   -\lambda(p-{3\over 2} p^2)q^2 - \sigma q+\mu\, ;
\label{dotq-run} \\
  \dot{p} &=& \lambda (p^2-p^3) q + \sigma(p-1)\, .   \label{dotp-run}
\end{eqnarray}
 \es
As always, the mean--field equation of motion for the reaction
coordinate $\bar q\approx \langle n\rangle$ is obtained by the
anzatz $p=1$ and takes the form
\begin{equation}
{d \bar q\over dt} = {\lambda\over 2}\,  \bar q^2-\sigma \bar q
+\mu\, .
                           \label{meanfield-run}
\end{equation}
According to  the mean--field equation there are two qualitatively
different scenarios of the system's evolution. They are
distinguished by the parameter
\begin{equation}
\delta^2 \equiv 1-\frac{2\lambda\mu}{\sigma^2}\, .
                              \label{delta}
\end{equation}
If $\delta^2<0$, the r.h.s. of Eq.~(\ref{meanfield-run}) is
strictly positive and the reaction coordinate always grows to
infinity. This is the scenario, where the population proliferates
indefinitely. Alternatively, for $\delta^2 >0$ the system possess
two stationary concentrations: $n_{\mp}=n_s(1\mp \delta)$, where
$n_s=\sigma/\lambda$. The point $\bar q = n_-$ is the stable one,
while $\bar q=n_+$ is unstable. In this case (the only one we
discuss hereafter), the mean--field predicts that for the range of
initial concentrations $0<n_0<n_+$ the system evolves towards the
stable population $n_-$. If the initial concentration exceeds
$n_+$ -- the system runs away and the population diverges.

\begin{figure}
\includegraphics[width=8cm]{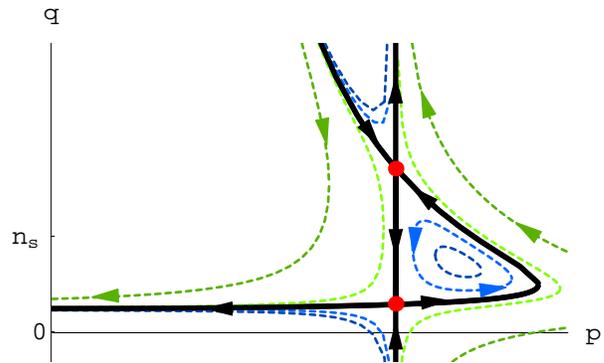}
\caption{The phase portrait of the run--away process,
Eq.~(\ref{master-run}). Thick  lines represent solution of $H(p,q)
= 0$, fat dots -- fixed points. $\delta^2 = 1/2$.}\label{run_away}
\end{figure}

If one goes beyond the mean--field treatment, however, one
realizes that the state $n_-$ is actually a {\em meta-stable} one.
To see this fact  and calculate the life-time of the meta-stable
state, it is convenient to draw the phase portrait,
Fig.~\ref{run_away}. It has two lines of zero energy: the
mean--field one, $p=1$, and the non--trivial line $\lambda p^2
q^2/2-\sigma q +\mu=0$. These two lines intersect at the
mean--field stable points $p=1\;;\; q=n_\mp$ and determine the
topology of the phase diagram. It is clear from the phase portrait
that the point $p=1\; ;\; q=n_-$ is not stable once motion with
$p\neq 1$ (non--mean--field) is allowed. More precisely, there is
a non--mean--field path that brings the system from the point
$q=n_-$ to the point $q=n_+$. Once the point $q=n_+$ is reached,
the system may continue to evolve according to the mean--field
towards indefinite population grow. Repeating calculations,
similar in spirit to calculations of the decay--time in section
\ref{s3}, one finds for the life--time of the meta--stable state,
$q=n_-$:
\begin{equation}
\tau\approx \sigma^{-1}\,\exp\{+S_0\}\, ,
                                 \label{lifetime}
\end{equation}
where $S_0$ is the classical action along the non--trivial line of
zero energy between points $(1,n_-)$ and $(1,n_+)$. Calculating
the integral, one finds $S_0=f(n_+)-f(n_-)$, where $f(x)\equiv
x-\sqrt{8\mu/\lambda}\;\arctan(x\sqrt{\lambda/2\mu})$.

Two limiting cases are of particular interest: (i) the "near
critical" system, $0<\delta^2\ll 1$ and (ii) the system with
almost no immigration, $\mu\to 0^+\; ;\; \delta\to 1^-$. In the
former case the two mean--field stationary points approach each
other, making the escape from the meta-stable state relatively
easy. Expanding the $f$-function up to the third (!) order, one
finds $S_0=2n_s\delta^3/3\ll n_s$. As expected, the action is
small and correspondingly the life--time is short (notice that the
quasi-classical picture applies as long as $S_0>1$). In the latter
case the two mean--field stationary points tend  to $n_-\to 0$,
and $n_+\to 2n_s$. If the immigration is absent, $\mu=0$, the
mean--field stable point, $n_-=0$, coincides with the empty state
of the system. The empty state is absolutely stable since no
fluctuations are possible. Naively one may expect that in this
limit the life--time of the meta-stable state (and thus $S_0$)
diverges. This is {\em not} the case, however. The calculation
shows: $S_0\to 2n_s$. As a result, even negligibly small
probability of immigration, $\mu$,  leads to a finite probability
of unlimited population expansion. (Strictly speaking, one also
needs to show that the pre-exponential factor does not go to zero
once $\mu\to 0$.)

We consider now a finite--dimensional generalization of this
population dynamics model.  The physics of the phenomena,
discussed here, is as follows: if a critically large cluster
"tunnels" into the run-away state, both diffusion and reaction
dynamics work to expand the cluster and flip the entire system
into the run-away mode. The situation is similar to  nucleation of
a critical domain in the super-cooled state of a system close to a
first order phase transition. To simplify the algebra we shall
consider only the case of the "near critical" system,
$0<\delta^2\ll 1$, where  the apparatus turns out to be rather
similar to that of the theory of the first order phase
transitions.

As discussed above, the finite--dimensional generalization of the
Hamiltonian is $H[p,q]=\int\! d^d x\, [H_0(p,q)+D\nabla p \nabla
q]$. For $\delta\ll 1$, it is convenient to make a change of
variables $(p,q)\to (\hat \varphi,\varphi)$, as $p=1+\hat \varphi$
and $q=n_s(1+\varphi)$, where $\varphi\sim \delta$, while $\hat
\varphi\sim \delta^2$. Substituting it into the reaction part of
the Hamiltonian, Eq.~(\ref{H-run}), and keeping terms up to
$\delta^4$, one obtains $H_0(\hat \varphi,\varphi)= \sigma
n_s[\hat \varphi(\delta^2 - \varphi^2)/2-\hat \varphi^2]$. As a
result, the $d$-dimensional action, Eq.(\ref{classaction-diff}),
for the conjugated fields $\hat \varphi(x,t)$ and $\varphi(x,t)$
takes the form:
\begin{equation}
  S = n_s\xi^d\int\limits_0^t\!\! dt\!\! \int\!\!  d^dx \left[
  \hat \varphi\left( \dot \varphi - \nabla^2 \varphi +
{\delta^2-\varphi^2\over 2}
  \right) - \hat \varphi^2\right],
                                              \label{classaction-c}
\end{equation}
where we have introduced the dimensionless time $\sigma t\to t$
and coordinate $x/\xi\to x$, where $\xi=\sqrt{D/\sigma}$. The
functional integration over the field $\hat \varphi$ should be
understood as running along the imaginary axis. The field theory
with the action Eq.~(\ref{classaction-c}) may be considered as
Martin--Sigia-Rose \cite{MSR} representation of the following
Langevin equation:
\begin{equation}
{\partial \varphi\over \partial t} =  \nabla^2 \varphi -{\partial
V\over
\partial \varphi} + \zeta(x,t)\, ,
                                        \label{Langevin}
\end{equation}
where  $\zeta(x,t)$ is a Gaussian noise with the correlator
\begin{equation}
\langle \zeta(x,t)\zeta(x',t')\rangle = \frac{2}{n_s\xi^d} \,
\delta(x-x')\delta(t-t')\,
                                         \label{noise}
\end{equation}
and the potential is $V(\varphi)=-\varphi^3/6+\delta^2
\varphi/2\;$. This potential has a meta-stable minimum at
$\varphi=-\delta$ and an unstable maximum at $\varphi=\delta$. The
barrier hight is $V(\delta)-V(-\delta) =2\delta^3/3$ and therefore
the life-time of the zero dimensional $(d=0)$ system is expected
to be given by the activation exponent (with $(n_s\xi^d)^{-1}$
playing the role of temperature) $\sim \exp\{n_s2\delta^3/3\}$, in
agreement with Eq~(\ref{lifetime}).

To discuss the life-time of the finite--dimensional system we
shall not use the Langevin approach, but rather return to the
action, Eq.~(\ref{classaction-c}), and write down the classical
equations of motion:
 \bs
\label{dot1-run}
\begin{eqnarray}
  \partial_t \varphi &=& \nabla^2 \varphi - {\partial V\over \partial \varphi}
+2\hat \varphi \, ; \label{dotq1-run} \\
  \partial_t \hat  \varphi &=& -\nabla^2 \hat \varphi + \hat \varphi\,
{\partial^2 V\over
\partial \varphi^2}\, .    \label{dotp1-run}
\end{eqnarray}
 \es
The energy density, corresponding to these two equations, is
defined as $E(x,t)=-\hat \varphi(\nabla^2 \varphi - \partial
V/\partial \varphi +\hat \varphi)$. The global energy,
$E=\int\!d^d x\, E(x,t)$, is, of course, conserved. However, in
the present case if $E(x,0)=0$ it keeps holding {\em locally} at
any time: $E(x,t)=0$. Indeed, the energy density vanishes if
either $\hat \varphi=0$, or $\hat \varphi= -\nabla^2 \varphi +
\partial V/\partial \varphi=2\hat \varphi -\partial_t \varphi$ and thus $\hat
\varphi=
\partial_t \varphi$, where we have employed Eq.~(\ref{dotq1-run}). It
is easy to check that in both cases Eq.~(\ref{dotp1-run}) is
satisfied automatically. Therefore the evolution with zero energy
density is described by either $\partial_t \varphi = \nabla^2
\varphi -
\partial V/\partial \varphi$, that is the mean--field equation, or by
$\partial_t \varphi =
-\nabla^2 \varphi + \partial V/ \partial \varphi$, that gives the
motion along the non-trivial line of zero energy.

Notice that the last equation happens to be the time-reversed
version of the mean--field \cite{IoffeSherrington}. If one starts,
thus, from the stationary solution $\varphi=-\delta$ and perturbs
it infinitesimally -- the perturbation  grows until it reaches the
stable configuration, satisfying
\begin{equation}
\nabla^2 \varphi - {\partial V\over \partial \varphi} =0\, .
                          \label{critdomain}
\end{equation}
The critical domain is given therefore by a localized solution of
Eq.~(\ref{critdomain}). Since the energy along the nucleation
dynamics is zero, the action to nucleate the critical domain is
given by $S_d=n_s\xi^d\int\!d^d x\int\! dt\, \hat \varphi
\,\partial_t \varphi= n_s\xi^d\int\!d^d x\int\! dt\, (-\nabla^2
\varphi +
\partial V/\partial \varphi)\, \partial_t \varphi$. Performing the time
integration in this expression, one finds for the action
\begin{equation}
S_d=n_s\xi^d\int\!d^d x\, \left({1\over 2} (\nabla \varphi_d)^2 +
V(\varphi_d)-V(-\delta)\right) ,
                          \label{Sd}
\end{equation}
where $\varphi_d=\varphi_d(x)$ is a stationary localized solution
of Eq.~(\ref{critdomain}), that is an extremum of the functional
(\ref{Sd}). As a result, the problem of the dynamical escape from
the meta-stable configuration is reduced to the static Landau
theory of the first order transitions. As far as we know, such
reduction is not a general statement, but rather is a consequence
of the assumption $\delta\ll 1$ and the resulting {\em local}
energy conservation, $E(x,t)=0$. In a general situation, one still
has to solve a considerably more complicated problem of  dynamic
equations (\ref{dot-run}) for $\varphi(x,t)$ and $\hat
\varphi(x,t)$.

>From the scaling analysis of Eq.~(\ref{critdomain}) one finds that
$\varphi\sim\delta$ in the core of the critical domain. Employing
this fact, one finds that the characteristic spatial scale of the
domain is given by $\delta^{-1/2}\gg 1$ (distance is measured in
units of $\xi=\sqrt{D/\sigma}\,$). Therefore the action cost to
create the critical domain is
\begin{equation}
S_d=c_d\,n_s\xi^d\,\delta^{3-d/2}\, ,
                          \label{Sd1}
\end{equation}
where $c_d$ is a numerical factor of order of one: $c_0=2/3\;;\;
c_1=24/5$. This result suggests that for $d>6$ the state with
finite population density $n=n_s(1-\delta)$ is stable, while for
$d<6$ the state is meta-stable. The concentration of critical
domains is given by $\xi^{-d}\exp\{-S_d\}$ and the typical
distance between them is $\xi\exp\{S_d/d\}$. They grow diffusively
until the entire system is flipped over to the run--away state in
time $\tau\sim \sigma^{-1}\exp\{2S_d/d\}$. The semiclassical
calculation is applicable as long as $S_d>1$ and therefore
$\delta$ is not too small. For very small $\delta$ the escape is
driven by the fluctuations rather than the semiclassical dynamics.

\section{Conclusions}
\label{conclusion}

The examples, considered above, are meant to illustrate the
general technique to calculate probability of rare events in
reaction--diffusion systems. The technique is based on the
existence of the many--body "quantum" Hamiltonian, which fully
encodes the microscopic Master equation. The very same Hamiltonian
in its second quantized representation serves as a starting point
for  field--theoretical treatments of dynamic phase transitions in
the reaction--diffusion system
\cite{OvchinnikovZeldovich78,ToussaintWilczek83,CardyGrassberger85,Hinrichsen,MarroDickman,Lee94,CardyTauber96,WOH98,Janssen01}.
For our present purposes we have deliberately chosen to work with
systems that are away from a possible continuous phase transition
point. Namely, we focus on the parts of the phase diagram, where
the mean--field considerations suggest a non--vanishing population
of particles (or at least transiently non--vanishing population).
In such cases the "quantum" fluctuations are small and one may
treat the underlying "quantum" dynamics in the semiclassical way.

We stress  that the  semiclassical treatment is {\em not}
equivalent to the mean--field one. The latter requires a very
special assumption about dynamics of the canonical momenta:
$p(x,t)=1$. This assumption may be justified as long as one is
interested in a typical system's behavior (even this is not
guaranteed if the system possess meta-stable states, as in our
last example). In such cases the problem is reduced to a partial
differential equation for the reaction coordinates, $q(x,t)$,
only. However, if questions about atypical, rare events are asked,
-- the mean--field assumption, $p(x,t)=1$, must be abandoned. As a
result, one has to deal with the canonical pair of the Hamilton
equations for reaction coordinates, $q(x,t)$, and momenta,
$p(x,t)$. The degree of deviation from the mean--field line,
$p=1$, is specified (through  proper initial and finite boundary
conditions) by the concrete sort of the rare event of interest.
Finally, the probability of the rare event is proportional to the
exponentiated action along the classical trajectory, satisfying
specified boundary conditions.

We found it especially useful to work with the phase portrait of
the corresponding dynamical system on the $(p,q)$ plane. The
emerging structures are pretty intuitive and can tell a great deal
about qualitative  behavior of the system even before any
calculations. The Hamiltonians underlying the Master equations of
reaction systems are typically {\em not} of the type traditionally
considered in the theory of dynamical systems. For example, they
usually can {\em not} be cast into the familiar form
$H(p,q)=p^2/2+A(q)p+V(q)$. On the other hand, they posses some
universal features, such as $H(1,q)=0$, or, if there is an empty
absorbing state,  $H(p,0)=0$, etc. These features dictate a
specific topology of the phase portrait. It would be extremely
interesting to explore this class of Hamiltonians from the point
of view of mathematical theory of dynamical systems \cite{Arnold}.
A question of particular interest is a possible exact
integrability of resulting Hamiltonian equations (especially in
$d=1$) \cite{Soliani}.

There are number of issues, that are not addressed in the present
paper and require further investigation. Let us mention some of
them. (i) Throughout the paper we have discussed the rare events
probability with the exponential accuracy. In some cases this is
not enough and one wants to know the pre-exponential factor rather
precisely.  This requires calculation of the fluctuation
determinant on top of the non--trivial classical trajectory. This
task is relatively straightforward for the $d=0$ systems, where it
may be addressed by writing down "quantum" corrections to the
Hamilton-Jacoby equation and treating them iteratively (in the way
it is usually done in the single--particle WKB method). For
extended systems the task is reduced to the spectral problem of a
certain matrix differential operator. At present we are not aware
of a general recipe to solve it. One may show, however, that on
any mean--field trajectory, $p(x,t)=1$, the fluctuation
determinant is equal to unity. The simplest way of doing it is to
use the discrete representation of the functional integral,
Eq.~(\ref{pathintegral}), and notice that the quadratic
fluctuation matrix has a triangular structure with unities on the
main diagonal (and, hence, unit determinant). Unfortunately, this
is not the case away from the mean--field, $p\neq 1$.

(ii) We have restricted ourselves to the systems with a single
sort of spices only. It is straightforward to generalize the
technique to any number of spices, $K$. The difficulty is that the
phase portrait becomes a $2K$--dimensional construction, which is
not easy to visualize. Correspondingly, the mean--field line
becomes $K$--dimensional hyperplane. Moreover some qualitatively
new physics may arise such as stable oscillatory limiting cycles
on the mean--field hyperplane. A paradigm of such behavior is a
Lotka--Volterra \cite{Lotka} system:
$A+B\stackrel{\lambda}{\longrightarrow} 2A$;
$A\stackrel{\sigma}{\longrightarrow} \emptyset $ and $B
\stackrel{\mu}{\longrightarrow} 2B$. An example of rare event may
be an "escape" from the periodic limiting cycle on the $A-B$
mean--field plane into the empty state in a finite size system.
Finding an optimal "reaction path" for such escape is not a
straightforward matter, however.

(iii) We have not treated long--range interactions and (local or
non-local) constraints. The simplest ("fermionic") constraint is
that of a maximum single occupancy of each lattice site. It was
shown recently that such constraint may be incorporated into the
"bosonic" formulation \cite{BOW00}, leading to a new class of the
interesting Hamiltonians. Studding rare event statistics for such
hard-core particles (by studding classical dynamics of the
corresponding Hamiltonians) is a very interesting subject.

(iv) There is a close resemblance between the formalism presented
here for essentially classical systems and the Keldysh technique
for non--equilibrium quantum statistics \cite{Kamenev02}. The
semiclassical solutions with $p\neq 1$, considered here,
correspond to saddle point configurations of the Keldysh action
with a different behavior on the forward and backward branches of
the time contour. Although examples of such saddle points were
considered in the literature \cite{Nazarov99,AtlandKamenev00}, it would
be interesting to learn more about possible applications of the
present technique for true quantum problems.

We are grateful to A. Elgart, Y. Gefen, A. Lopatin and K. Matveev
for useful conversations. A.~K. is A.~P.~Sloan fellow.

\begin{appendix}
\section{Operator technique}
\label{appendix}

We give here a brief account of the operator technique
\cite{Doi76,Peliti84,MattisGlasser98,Cardy_html} for completeness.
Define the ket--vector $|n\rangle$ as the microscopic state with
$n$ particles. Let us also define vector
\begin{equation}
|\Psi(t)\rangle \equiv \sum\limits_{n=0}^{\infty} P_n(t)\,
|n\rangle \, .
                                              \label{Apsi}
\end{equation}
Notice, that the weight, $P_n$, is probability rather  than  the
amplitude. It is convenient to introduce the creation and
annihilation operators with the following properties:
 \bs
 \label{Aoper}
 \begin{eqnarray}
a^{\dagger}|n\rangle  &=&  |n+1 \rangle \, ;    \label{Aa} \\
a|n\rangle  &=& n |n-1 \rangle   \, .  \label{Aadagger}
\end{eqnarray}
 \es
As a byproduct, one has $a|0\rangle = 0$. One may immediately
check that such operators are ``bosonic'':
\begin{equation}
[a,a^{\dagger}] = 1 \, .
                                            \label{Acommutation}
\end{equation}
As for any pair of operators satisfying Eq.~(\ref{Acommutation})
one may prove the identity
\begin{equation}
e^a f(a,a^{\dagger}) =  f(a,a^{\dagger} + 1) e^a \, ,
                                            \label{Aidentity}
\end{equation}
where $f$ is an arbitrary operator--value function. In these
notations the whole set of the Master equations may be recast into
single "imaginary time" Schr\"odinger equation
\begin{equation}
\frac{d}{dt}\, |\Psi(t)\rangle = -\hat H |\Psi(t)\rangle,
                                              \label{Ashrodinger}
\end{equation}
where $\hat H$ is the ``Hamiltonian'' operator.  One may check
that the Hamiltonian of the binary annihilation process,
Eq.~(\ref{master}), has the form
\begin{equation}
\hat H = {\lambda \over 2}  \big( (a^{\dagger})^2  - 1 \big) a^2
\, ,
                                            \label{AH}
\end{equation}
where the first term in brackets on the r.h.s. is the ``out''  and
the second one is the ``in'' term.

One may solve formally the Schr\"odinger equation and write
$|\Psi(t) \rangle = \exp\{-\hat H(a^{\dagger}, a) t\} |\Psi(0)
\rangle$. An initial state, $|\Psi(0) \rangle$, is specified as
e.g. $|\Psi(0) \rangle = e^{-n_0(a^{\dagger} - 1)} |0\rangle$ for
the Poisson initial distribution, or $|\Psi(0) \rangle =
\left(a^{\dagger}\right)^{n_0} |0\rangle$ for the fixed particle
number. The generating function Eq.~(\ref{generating}) is given by
\begin{equation}
G(p,t) = \langle 0| e^{p a}\, e^{-\hat H(a^{\dagger}, a) t}
|\Psi(0) \rangle \, .
                                            \label{Agen_func}
\end{equation}
The normalization, $G(1,t=0)=1$,  is guaranteed by the identity
$\langle 0|e^a| n\rangle = 1$ for any $n$ (this fact may be
checked using Eq.~(\ref{Aidentity})) and the fact $\sum_n
P_n(0)=1$. The normalization is kept intact at any time if
$\langle 0|e^a \hat H(a^{\dagger}, a) =0 $. Since the coherent
state $\langle 0|e^a$ is an eigenstate of the creation operator,
$\langle 0|e^a a^{\dagger} = \langle 0|e^a$, one arrives at the
conclusion that any legitimate Hamiltonian must obey
\begin{equation}
\hat H(a^{\dagger} = 1, a) = 0 \, .
                                            \label{Anorm1}
\end{equation}
E.g. the Hamiltonian of the binary annihilation, Eq.~(\ref{AH}),
indeed satisfies this condition.

One may employ now the standard bosonic coherent state technique
to write the generating function, Eq.~(\ref{Agen_func}), as the
functional integral. The result coincides identically with the
Eq.~(\ref{pathintegral}) of the main text. One notices, thus, the
formal correspondence between the operators $a^\dagger$ and $a$
and operators $\hat p$ and $\hat q$ correspondingly.

\end{appendix}




\begin{thebibliography}{xxx}

\bibitem{Kampen}
N. G. Van Kampen,
\newblock {\it Stochastic Processes in Physics and Chemistry},
\newblock {North-Holland}, (1981).

\bibitem{Gardiner}
C W Gardiner,
\newblock {\it A Handbook of Stochastic Methods},
\newblock {Springer, Berlin Heidelberg}, (1998).

\bibitem{ZGB86}
R.~M. Ziff, E.~Gulari, and Y.~Barshad,
\newblock Kinetic phase transitions in an irreversible surface-reaction model,
\newblock {Phys. Rev. Lett.} {\bf 56}, 2553--2556 (1986).

\bibitem{Murray} See, e.g.,  J. D.  Murray,{\it  Mathematical   Biology}
(Springer-Verlag, N.Y., 1993) and references therein.

\bibitem{Albano94a}
E.~V. Albano,
\newblock Critical behaviour of a forest fire model with immune trees,
\newblock {J. Phys.} {\bf A 27}, L881--L886 (1994).

\bibitem{Berry}
H. Berry,
\newblock {Phys. Rev. E} {\bf 67}, 031907 (2003).

\bibitem{Liggett85}
T.~M. Liggett,
\newblock {\em Interacting particle systems},
\newblock Springer, Berlin, 1985.

\bibitem{Mollison77}
D.~Mollison,
\newblock Spatial contact models for ecological and epidemic spread,
\newblock J. Roy. Stat. Soc. {\bf B 39}, 283 (1977).

\bibitem{BouchaudGeorges90}
J.-P. Bouchaud and A.~Georges,
\newblock Anomalous diffusion in disordered media: statistical mechanics,
  models and physical applications,
\newblock Phys. Rep. {\bf 195}, 127--293 (1990).

\bibitem{Okubo} A. Okubo, {\it Diffusion and Ecological Problems:
Mathematical  Models}  (Springer-Verlag, N.Y., 1980).

\bibitem{OvchinnikovZeldovich78} A.~A.~Ovchinnikov and Ya.~B.~Zeldovich, Chem.
Phys. {\bf 28}, 215 (1978).

\bibitem{ToussaintWilczek83} D. Toussaint and F. Wilczek, J. Chem. Phys. {\bf
78}, 2642 (1983).

\bibitem{CardyGrassberger85}
J.~L. Cardy and P.~Grassberger,
\newblock Epidemic models and percolation,
\newblock {J. Phys.} {\bf A 18}, L267--L271 (1985).

\bibitem{Hinrichsen}
\newblock H.~Hinrichsen,  Non-equilibrium critical phenomena and phase transitions into absorbing states,
Adv. Phys., {\bf 49}, 815 (2000).


\bibitem{MarroDickman}
J. Marro and R. Dickman,
\newblock {\it Nonequilibrium Phase Transitions in Lattice Models}
\newblock {Cambridge University  Press, 1999}.




\bibitem{Lee94}
B.~P. Lee,
\newblock Renormalization group calculation for the reaction kA$\rightarrow 0$,
\newblock {J. Phys.} {\bf A 27}, 2633--2652 (1994).

\bibitem{CardyTauber96}
J.Cardy and U.~C. T{\"a}uber, Phys. Rev Lett. {\bf  77}, 4780
(1996).

\bibitem{WOH98}
F.~van Wijland, K.~Oerding, and H.~J. Hilhorst,
\newblock Wilson renormalization of a reaction-diffusion process,
\newblock Physica {\bf A 251}, 179--201 (1998).

\bibitem{Janssen01}
V. Becker and H. K. Janssen,
 Current-current correlation function in a driven diffusive system with nonconserving
 noise,
Phys. Rev. E {\bf 50}, 1114 (1994);\\ H. K. Janssen and K. Oerding,
 Renormalized field theory and particle density profile in driven diffusive systems with open
 boundaries,
Phys. Rev. E {\bf 53}, 4544-4554 (1996).


\bibitem{CardySugar80}
J.~L. Cardy and R.~L. Sugar,
\newblock Directed percolation and Reggeon field theory,
\newblock {J. Phys.} {\bf A 13}, L423--L427 (1980).

\bibitem{Grassberger97}
P.~Grassberger,
\newblock Directed percolation: results and open problems,
\newblock in {\em Nonlinearities in complex systems, proceedings of the 1995
  Shimla conference on complex systems}, edited by S.~{Puri {\it et al.}},
  Narosa Publishing, New Dehli, 1997.

\bibitem{Janssen81}
H.K. Janssen,
\newblock On the nonequilibrium phase transition in reaction-diffusion
systems with an absorbing stationary state,
\newblock {Zeitschrift fur Physik B}, {\bf 42}, 151--4 (1981).

\bibitem{Doi76}
M.~Doi,
\newblock Stochastic theory of diffusion-controlled reactions,
\newblock {J. Phys.} {\bf A 9}, 1479--1495 (1976).

\bibitem{Peliti84}
L.~Peliti,
\newblock Path integral approach to birth-death processes on a lattice,
\newblock {Journal de Physique} {\bf 46}, 1469--1483 (1984).

\bibitem{MattisGlasser98}
D.~C. Mattis and M.~L. Glasser,
\newblock The uses of quantum field theory in diffusion-limited reactions,
\newblock {Rev. Mod. Phys.} {\bf 70}, 979--1001 (1998).



\bibitem{Jordan}
S. Pilgram, A. N. Jordan, E. V. Sukhorukov, and M.  Buttiker,
\newblock Stochastic Path Integral Formulation of Full Counting
Statistics,
\newblock {Phys. Rev. Lett.} {\bf 90}, 206801 (2003);
A. N. Jordan, E. V. Sukhorukov, and S. Pilgram,
\newblock Fluctuation Statistics in Networks: a Stochastic Path Integral
Approach,
\newblock {cond-mat/0401650}.


\bibitem{Bagrets}D. A. Bagrets and Yu. V. Nazarov, Full counting statistics of charge
transfer in Coulomb blockade systems, Phys. Rev. B {\bf 67},
085316 (2003).

\bibitem{footJ} Canonical pair of variables of
Ref.~[\onlinecite{Jordan}] differs from ours by the canonical
transformation $p=e^{i\Lambda}$, $q=e^{-i\Lambda}Q$. We are
gratefull to the authors of Ref.~[\onlinecite{Jordan}] for
clarification of this point.


\bibitem{Zinn-Justin} See, e.g.,  J. Zinn-Justin,{\it  Quantum Field Theory
  and Critical Phenomena}
(Oxford Press; 4th edition, 2002).

\bibitem{Lindenberg03}
C. Escudero, J. Buceta, F.J. de la Rubia, and K. Lindenberg,
\newblock Extinction in Population Dynamics,
\newblock {cond-mat/0309568}.

\bibitem{MSR}
P. C. Martin, E. D. Siggia and H. A. Rose
\newblock {Statistical dynamics of classical systems},
\newblock {Phys. Rev. A, Gen. Phys.} {\bf 8} 423-37, (1973).

\bibitem{IoffeSherrington}
L. B. Ioffe and D. Sherrington,
\newblock Distribution of barriers in spin glasses,
\newblock {Phys. Rev. B} {\bf 57}, 7666, (1998).


\bibitem{Arnold}
V.I. Arnold,
\newblock {\it Mathematical Methods of Classical Mechanics},
\newblock (2nd ed., Springer-Verlag, N.Y., 1989).

\bibitem{Soliani}
E. Alfinito, V. Grassi, R. A. Leo, G. Profilo, G. Soliani,
\newblock Equations of the reaction-diffusion type with a loop algebra
structure,
\newblock {Inv. Prob.} {\bf 14}, 1387-1401 (1998).

\bibitem{Lotka}A. J. Lotka, Proc. Natl. Acad. Sci. USA {\bf 6}, 410
(1920);
V. Volterra, {\it Lecon sur la Theorie Mathematique de la Lutte
pour le via } (Gauthier-Villars, Paris, 1931).

\bibitem{BOW00}
V.~Brunel, K.~Oerding, and F.~van Wijland,
\newblock Fermionic field theory for directed percolation in (1+1) dimensions,
\newblock {J. Phys.} {\bf A 33}, 1085--1097 (2000).

\bibitem{Kamenev02}
A. Kamenev,
\newblock Keldysh and Doi-Peliti Techniques for out-of-Equilibrium
Systems, in {\em "Strongly Correlated Fermions and Bosons in
Low-Dimensional Disordered Systems"},
\newblock {I. V. Lerner, et. al. editors, \it Kluwer Academic Publishers,
Dordrecht, Boston, London} 313-340 (2002).

\bibitem{Nazarov99}
Yu. V. Nazarov,
\newblock Universalities of weak localization,
\newblock {Ann. Phys. (Leipzig)} {\bf 8}, 507 (1999).

\bibitem{AtlandKamenev00}
A. Altland and A. Kamenev,
\newblock Wigner--Dyson Statistics from the Keldysh $\sigma$-Model,
\newblock {Phys. Rev. Lett.}, {\bf 85}, 5615--18 (2000).


\bibitem{Cardy_html} J.~Cardy,
http://www-thphys.physics.ox.ac.uk\\/users/JohnCardy/home.html




%
%

\end{thebibliography}
\end{document}